\documentclass[preprint,aps,prl,onecolumn,showpacs,preprintnumbers,amsmath,amssymb]{revtex4}

\usepackage{graphicx}
\usepackage{dcolumn}
\usepackage{bm}
\newcommand{\theq}{\theta_{\text{eq}}}
\newcommand{\vct}[1]{\mathbf{#1}}

\begin{document}


\title{Size dependent motion of nanodroplets on chemical steps}

\author{A.  Moosavi} 
\affiliation{Max-Planck-Institut f\"ur Metallforschung, Heisenbergstr. 3,
D-70569 Stuttgart, Germany,} 
\affiliation{Institut f\"ur Theoretische und Angewandte Physik,
Universit\"at Stuttgart, Pfaffenwaldring 57, D-70569
Stuttgart, Germany}

\author{M. Rauscher}
\affiliation{Max-Planck-Institut f\"ur Metallforschung,
Heisenbergstr. 3, D-70569 Stuttgart, Germany,}
\affiliation{Institut f\"ur Theoretische und Angewandte Physik,
Universit\"at Stuttgart, Pfaffenwaldring 57, D-70569
Stuttgart, Germany}

\author{S. Dietrich}
\affiliation{Max-Planck-Institut f\"ur Metallforschung,
Heisenbergstr. 3, D-70569 Stuttgart, Germany,}
\affiliation{Institut f\"ur Theoretische und Angewandte Physik,
Universit\"at Stuttgart, Pfaffenwaldring 57, D-70569
Stuttgart, Germany}

\date{\today}

\begin{abstract}

Nanodroplets on chemically structured substrates move under the action 
of disjoining pressure induced forces. A detailed analysis of them shows that even in the absence of 
long-ranged lateral variations of the effective interface potential, already the fact, 
that due their small size nano-droplets do not sample the disjoining pressure at all distances from the
substrate, can lead to droplet motion towards the less wettable part of the
substrate, i.e., in the direction opposite to the one expected on the basis of
macroscopic wettability considerations.

\end{abstract}


\maketitle
Chemically structured surfaces with their rather complex wetting properties 
\cite{dietrich99} are not only
ubiquitous in nature (e.g., beatles use them for harvesting drinking water out of
fog \cite{parker}), they have also found important applications in a wide
variety of processes ranging from ink-jet printing \cite{Wang:2004} and condensation heat transfer
\cite{Daniel:1992} to the so-called lab-on-a-chip concept for chemical analysis
or biotechnology which, e.g., allows one to handle minute amounts of liquid
containing DNA or proteins \cite{Karniadakis:2005, Dietrich:2005}).
In view of the richness of the corresponding wetting phenomena and in view of
their importance, they have been intensively studied both experimentally
\cite{Ondarcuhu:1991,Chaudhury:1992,Leopoldes:2005,Checco:2006} and theoretically
\cite{Greenspan:1978,Raphael:1988,Schwartz:1998,Bauer:1999,Bauer:1999b,Yaneva:2004,Springer:2005,Thiele:2006,Cieplak:2006,Yochelis:2007,Moosavi:2008}.

All the above mentioned applications of wettability patterns
rely on the fact that the fluid placed on such a substrate will actually move
towards the more wettable part \cite{parker,Daniel:1992,Karniadakis:2005,
Dietrich:2005}.  However, recently 
we could show, that nanodroplets near chemical steps can move in
the opposite direction
\cite{Moosavi:2008}. Due to the long
range of dispersion forces of the van der Waals type, droplets
residing on one side of the step perceive the different character of the other
side even at finite distances from the step and move accordingly. Since the
wetting properties (i.e., the equilibrium contact angles $\theq$) are determined
by the interplay between both the long- \textit{and} short-ranged interactions, the force
resulting only from the long-ranged part is not necessarily directed towards the
more wettable side \cite{Moosavi:2008}\/. 

In the present study we show that, in contrast to a droplet only \textit{near} a
chemical step, the direction of motion of a droplet positioned
\textit{on top of} a chemical step can depend on the size of the
droplet and that, depending on the substrate characteristics, sufficiently small
droplets move towards the less wettable side. For flat droplets this effect is robust, i.e.,
independent of the detailed properties of the chemical step (in contrast to the
phenomena involving the approach towards chemical steps \cite{Moosavi:2008}) and
it can be stronger as compared with this latter approach.
This is true in particular for
substrates the chemical patterns of which are generated by modifying the
\textit{short}-ranged part of the liquid-substrate interaction potential, as it is the
case for many fabrication techniques (see, e.g., Ref.~\cite{Checco:2006})\/. We
strengthen this prediction by mesoscopic hydrodynamic calculations for two-dimensional
droplets (i.e., filaments) and demonstrate its experimental relevance for
polystyrene nanodroplets on silicon oxide, a system for which the actual material
parameters are available.

\begin{figure}[!tb]
\center
\includegraphics[width=0.9\linewidth]{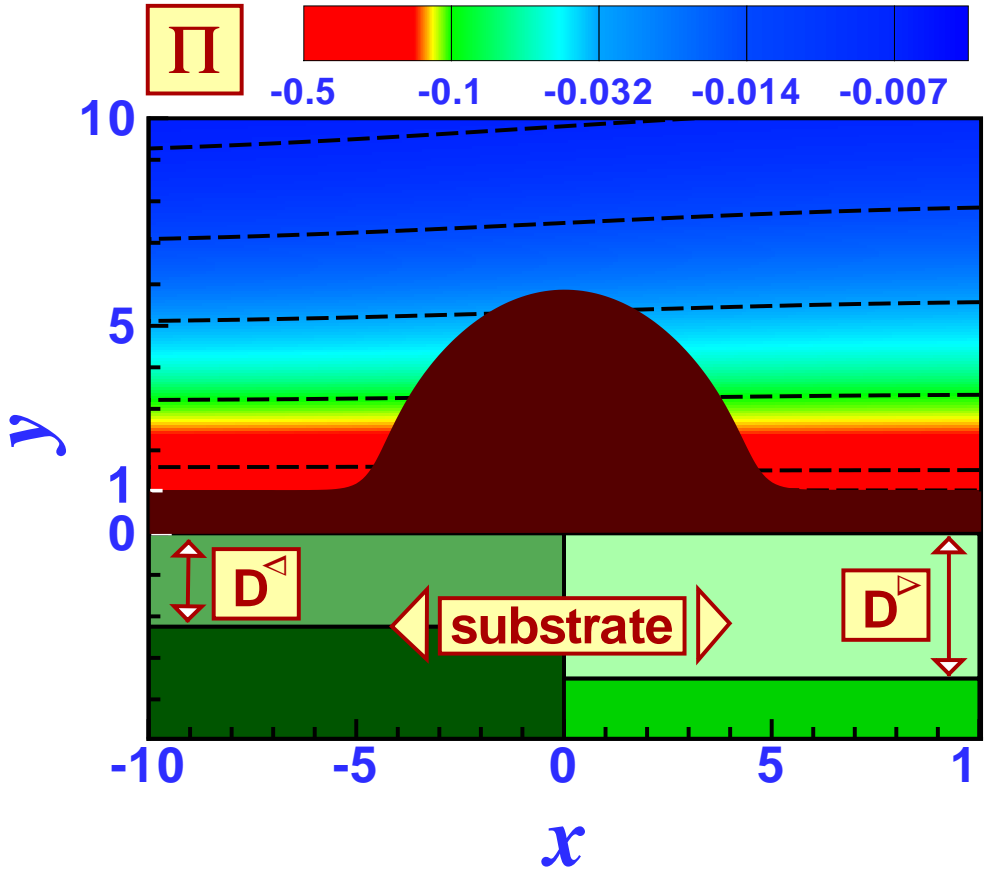}
\caption{A nanodroplet in contact with its corresponding equilibrium wetting film
and located on top of a chemical step between two materials with coatings of
different type and thickness. The contour and false color plot shows the
vertically and laterally varying DJP for the example of two
quarter substrates with different chemical composition but giving rise to the same
equilibrium contact angle  $\theta_{eq}^{\lhd}=\theta_{eq}^{\rhd}=90^\circ$ (on the corresponding laterally homogeneous substrates) formed
with the substrate surface. Due to the small size of the nanodroplet, its contact angles deviate from $\theta^{\lhd}_{eq}$ and $\theta^{\rhd}_{eq}$, respectively. The corresponding parameter values are $C^{\lhd}=1$,
$B^\lhd=7$, $D^\lhd=1.227$, and $C^{\rhd}=1$, $B^\rhd=14$, $D^\rhd=2.251$\/;
lengths and the DJP are measured in units of $b^{\lhd}$ and
$\sigma/b^{\lhd}$, respectively; see the main text for definitions.
}
\label{fig1}
\end{figure} 

We consider a droplet centered on a chemical step between two materials with
coatings of different thickness and nature as illustrated in Fig.~\ref{fig1}\/.
As in Refs.~\cite{Moosavi:2006,Moosavi:2008}, we calculate the
disjoining pressure (DJP)
by integrating pair potentials over the substrate. In addition, in order to
demonstrate the robustness of the aforementioned effect, following
Refs.~\cite{Schwartz:1998,Thiele:2006} we also use an approximate DJP which on
each side of the chemical step equals the corresponding one of a laterally
homogeneous substrate so that its lateral variation reduces to a discontinuity at
$x=0$.

A model chemical step as illustrated in Fig.~\ref{fig1} is
composed of two different quarter spaces (edges) with 
each of them coated on its upper side with a different material. We model the
intermolecular interactions between the liquid and the substrate as
well as among the fluid particles 
via pairwise additive Lennard-Jones type intermolecular pair potentials,
i.e., 
$V^{\lhd/\rhd}_{\alpha \beta}(r)={M^{\lhd/\rhd}_{\alpha
\beta}}/{r^{12}}-{N^{\lhd/\rhd}_{\alpha \beta}}/{r^6}$, where
$M^{\lhd/\rhd}_{\alpha \beta}$ and $N^{\lhd/\rhd}_{\alpha \beta}$
are material parameters for the left ($\lhd$) and right ($\rhd$) half of the
substrate, respectively, and $\alpha$ and $\beta$ relate to either liquid
($l$), edge ($e$), or coating ($c$) particles.
Following Ref.~\cite{Robbins:1991} the DJP $\Pi(x,y)$ of such a  
configuration can be calculated by integrating the liquid-substrate
interactions over the substrate and by subtracting the integral of the
liquid-liquid interaction potential over the volume occupied by the
substrate. 
For the contribution of the left $e$dge without surface coating occupying
$\Omega_e^{\lhd}=\left\{\vct{r}\in\mathbb{R}^3| x\le0 \wedge y \le0\right\}$ one obtains 
\begin{equation}
\Pi_e^{\lhd}(x,y) = \Delta M_{e}^{\lhd} \,I_{12}(x,y) - 
\Delta N_{e}^{\lhd} \,I_6(x,y),
\end{equation}
with $I_n(x,y) = \int_{\Omega_e^\lhd} \,|\vct{r}-\vct{r}'|^{-n}
\,d^3r'$,
$\Delta M_e^{\lhd/\rhd}= \rho_l^2 M_{ll}-\rho_l\rho^{\lhd/\rhd}_e M_{el}^{\lhd/\rhd}$,
and $\Delta N_e^{\lhd/\rhd}=\rho_l^2
N_{ll}-\rho_l\rho^{\lhd/\rhd}_e N_{el}^{\lhd/\rhd}$; $\rho_l$  and
$\rho^{\lhd/\rhd}_e$ are the 
number densities of the liquid and the edge, respectively. 
The integrals  $I_{12}(x,y)$ and $I_{6}(x,y)$ 
can be calculated analytically and can be used as suitable building blocks to
obtain the contribution of the $c$oating layer of thickness $D^{\lhd}$ and of the
right edge and its upper coating layer of thickness $D^{\rhd}$. With these,
the DJP of the chemical step is given by  
\begin{multline}
\label{fulldjp}
\Pi(x,y)=  \Pi_{ce}^{\lhd} + \Pi_{ce}^{\rhd}=\\
  \Delta M_c^{\lhd}\,\left[I_{12}(x,y)-I_{12}(x,y+D^{\lhd})\right]
\\
- \Delta N_c^{\lhd}\,\left[I_{6}(x,y)-I_{6}(x,y+D^{\lhd})\right] \\
+ \Delta M_e^{\lhd}\,I_{12}(x,y+D^{\lhd}) 
- \Delta N_e^{\lhd}I_{6}(x,y+D^{\lhd}) \\
 + \Delta M_c^{\rhd}\,\left[I_{12}(-x,y)-I_{12}(-x,y+D^{\rhd})\right]
\\
- \Delta N_c^{\rhd}\,\left[I_{6}(-x,y)-I_{6}(-x,y+D^{\rhd})\right]\\
+ \Delta M_e^{\rhd}\,I_{12}(-x,y+D^{\rhd})
- \Delta N_e^{\rhd}I_{6}(-x,y+D^{\rhd}),
\end{multline}
with $\Delta M_c^{\lhd/\rhd}= 
\rho_l^2 M_{ll}-\rho_l\rho^{\lhd/\rhd}_c
M_{cl}^{\lhd/\rhd}$ and $\Delta N_c^{\lhd/\rhd}=\rho_l^2
N_{ll}-\rho_l\rho^{\lhd/\rhd}_c N_{cl}^{\lhd/\rhd}$\/. 
For $x\to-\infty$ one has $I_{12}\to \pi/(45\,y^{9})$ and
$I_6\to \pi/(6\,y^{3})$ so that far from the chemical step the DJP
reduces to  $\Pi_{ch}^{\lhd}$ and
$\Pi_{ch}^{\rhd}$ of a $c$oated laterally $h$omogeneous substrate on
the left and right hand side, respectively.
For small distances from the substrate surface, i.e., for
$y\to 0$, the leading terms ($\sim \Delta M_c^{\lhd/\rhd}\,\pi/(45\,y^{9})$) 
in Eq.~\eqref{fulldjp} are the terms
$\Delta M_c^{\lhd}\,I_{12}(x,y)$ for $x<0$ and $\Delta
M_c^{\rhd}\,I_{12}(-x,y)$ for $x>0$, respectively. In order to
have a wetting film of finite thickness $b^{\lhd/\rhd}$, the
disjoining pressure has to become very large at small distances
from the substrate. Therefore, $\Delta M_c^{\lhd/\rhd}\ge0$ is a
necessary condition for the occurrence of an equilibrium wetting
film.
But the other amplitudes can be positive or 
negative. In order to simplify the presentation 
we only consider the case $\Delta N_c^{\lhd/\rhd}>0$ as the relevant one 
for the experimental systems studied in Refs.~\cite{Becker:2003,Seemann:2001,Seemann:2001b}\/. 
Also by limiting ourself to coating layers thick compared with the
wetting layer (see below) and to 
$|\Delta M_e^{\lhd/\rhd}|\ll\Delta M_c^{\lhd/\rhd}$, we neglect terms $\propto \Delta M_e^{\lhd/\rhd}$ \cite{Seemann:2001,Seemann:2001b}.
For this system 
we introduce dimensionless quantities (marked by a star) such that 
lengths are measured in units of $b^{\lhd}={[2\Delta M_c^{\lhd}
/(15\,\Delta N_c^{\lhd}|)}^{1/6}$, which is the equilibrium wetting film
thickness on a homogeneous flat substrate with the material of the left coating, and the DJP is measured in units of 
$\sigma/b^{\lhd}$ where $\sigma$ is the liquid-vapor surface tension as an independent parameter.
Thus the dimensionless DJP 
far from the edge $(x\rightarrow -/+\infty)$ has
the form
\begin{multline}
\label{dps}
\Pi^{*\lhd/\rhd}_{ch}(y_*)={C^{\lhd/\rhd}}{\Bigg[}\frac{{q^{\lhd/\rhd}}^9}{{y_*}^9} 
-\frac{{q^{\lhd/\rhd}}^9}{(y_*+D_*^{\lhd/\rhd})^9}\\
-\frac{{q^{\lhd/\rhd}}^3}{{y_*}^3}+\frac{{q^{\lhd/\rhd}}^3}{{(y_*+D_*^{\lhd/\rhd})}^3}-\frac{{q^{\lhd/\rhd}}^3B^{\lhd/\rhd}}{{(y_*+D_*
^{\lhd/\rhd})}^3}{\Bigg]},
\end{multline}
with $C^{\lhd/\rhd}=A^{\lhd/\rhd}\,b^{\lhd}/\sigma$,
$q^{\lhd/\rhd}=b^{\lhd/\rhd}/b^{\lhd}$ (i.e., $q^{\lhd}=1$), and 
$B^{\lhd/\rhd}=\Delta N_e^{\lhd/\rhd}/\Delta N_c^{\lhd/\rhd}$\/. The fully inhomogeneous and 
dimensionless DJP is obtained from Eq.~\eqref{fulldjp} by
replacing $\Delta M_c^{\lhd/\rhd}$ by
$45\,{q^{\lhd/\rhd}}^9\,C^{\lhd/\rhd}/\pi$, $\Delta
N_c^{\lhd/\rhd}$ by $6\,{q^{\lhd/\rhd}}^3\,C^{\lhd/\rhd}/\pi$,
and $\Delta N_e^{\lhd/\rhd}$ by
$6\,{q^{\lhd/\rhd}}^3\,C^{\lhd/\rhd}\,B^{\lhd/\rhd}/\pi$\/.
In order to avoid a clumsy notation in the following we drop the stars.
The macroscopic equilibrium contact angles
$\theta^{\lhd/\rhd}_{eq}$ far from the step are given by 
\cite{Dietrich:1988}
\begin{equation}
\label{cangle}
\cos\theta_{eq}^{\lhd/\rhd}=1+\int_{y_0^{\lhd/\rhd}}^\infty
\Pi_{ch}^{\lhd/\rhd}(y)\,dy.
\end{equation}
The wetting film thickness $y_0^{\lhd/\rhd}$ minimizes the effective interface potential $\omega^{\lhd/\rhd}(y)=
\int_y^\infty\Pi_{ch}^{\lhd/\rhd}(y')\,dy'$ 
and both $\theta^{\lhd/\rhd}$ and $y_0^{\lhd/\rhd}$ can be expressed in terms of $C^{\lhd/\rhd}$, $B^{\lhd/\rhd}$, and $D^{\lhd/\rhd} $\cite{Dietrich:1988,Moosavi:2006}\/. 

\begin{figure}[!tb]
\center
\includegraphics[width=0.9\linewidth]{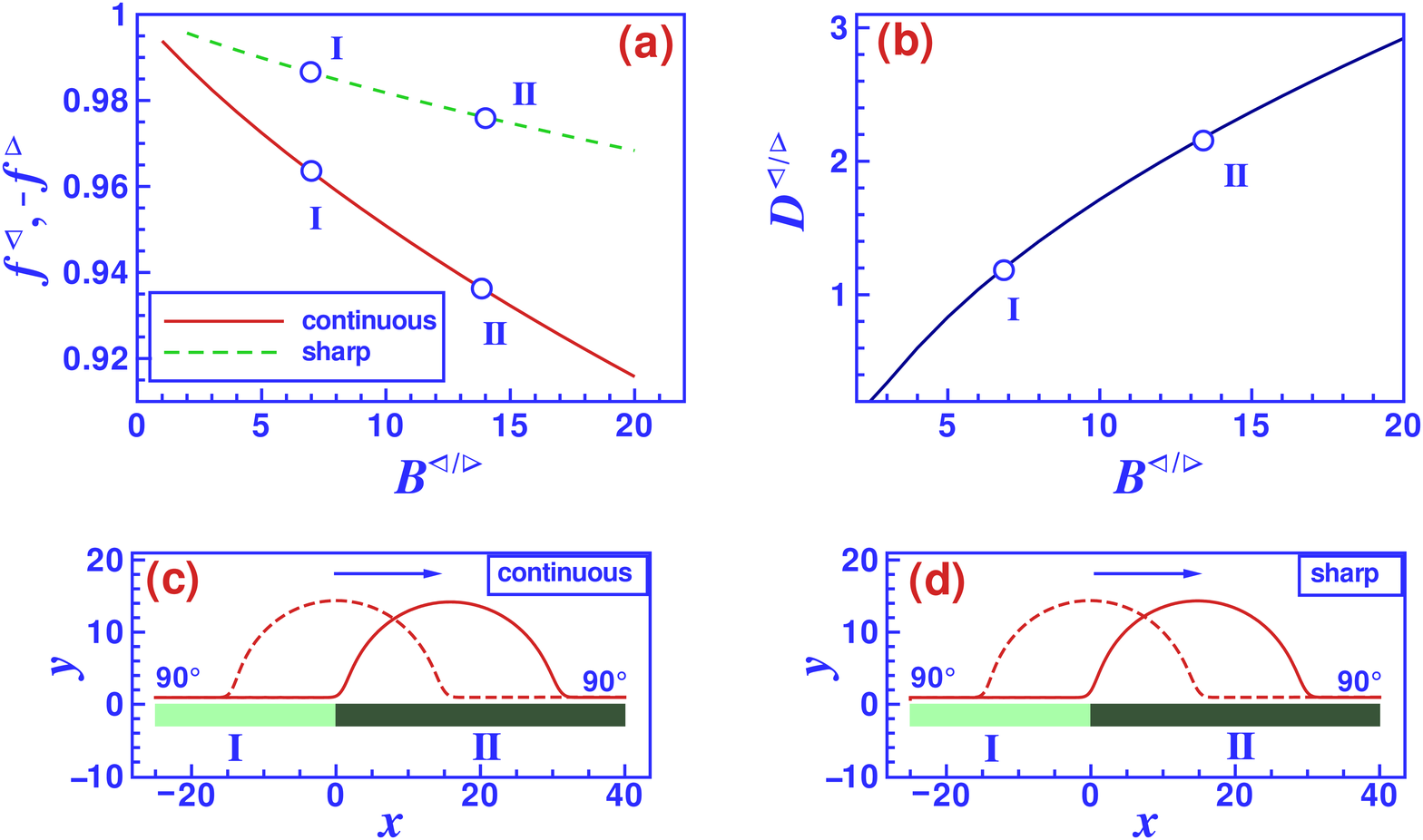}
\caption{(a): The continuous and the sharp DJP induced lateral
force $f^{\lhd/\rhd}$
originating from the left and the right part of the substrate,
respectively, on a droplet of height $a$ and basal width $2w$ with $a=w=15$ centered on the chemical step
as function of $B^{\lhd/\rhd}$ which is varied together with $D^{\lhd/\rhd}$ such that $\theta^{\lhd}_{eq}=\theta^{\rhd}_{eq}=90^\circ$ (b). If the the left (right) substrate is chosen to correspond to I (II), $f=f^{\lhd}+f^{\rhd}>0$. The force per unit length is measured in units of $\sigma$\/. 
(c) and (d): Stokes dynamics of the same droplet on a chemical step
between two substrates characterized by the materials parameters
corresponding to points I ($C^{\lhd}=1$, $q^{\lhd}=1$, $B^{\lhd}=7$, $D^{\lhd}=1.227$) and II ($C^{\rhd}=1$,
$q^{\rhd}=1$, $B^{\rhd}=14$, $D^{\rhd}=2.251$) in (a) and (b), respectively, for the
continuous (c) and the sharp (d) DJP. The dashed
line corresponds to the shape-relaxed initial state at time $t=70$ and the
solid lines to $t=5000$ (c) and $t=10500$ (d); times are given in units of
$\mu b^{\lhd}/(C^{\lhd}\sigma)$. The droplets move to the right
even though the equilibrium contact angles on both sides are
equal.}
\label{fig2}
\end{figure} 

\begin{figure}[!tb]
\center
\includegraphics[width=0.9\linewidth]{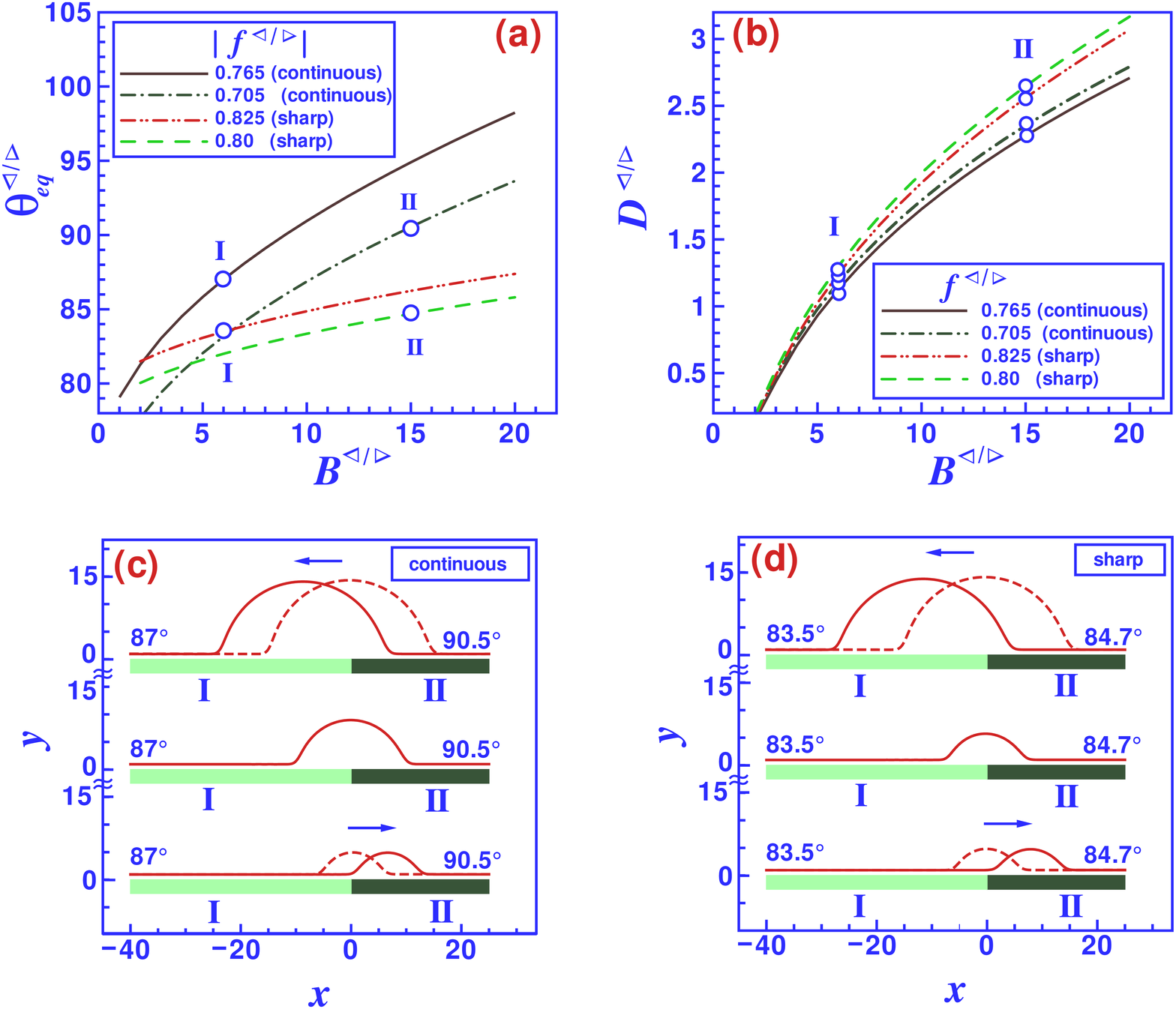}
\caption{(a), (b): Values of the parameters $B^{\lhd/\rhd}$ and $D^{\lhd/\rhd}$
and of the contact angle $\theta^{\lhd/\rhd}_{eq}$ for which the  absolute value of the lateral force
$|f^{\lhd/\rhd}|$ (originating from the left/right hand side of the substrate
on a droplet of height $a$ and width $2w$ with $a=w=5$ centered on the chemical step) is equal
to 0.705 or 0.765 (continuous) and 0.8 or 0.825 (sharp) in units
of $\sigma$\/. (c) and (d): Results of the Stokes dynamics for the
continuous (c) and sharp (d) DJP and for the parameter sets I
(continuous: $B^{\lhd}=6$, $D^{\lhd}=1.1246$; sharp: $B^{\lhd}=6$,
$D^{\lhd}=1.2414$) and II (continuous: $B^{\rhd}=15$,
$D^{\rhd}=2.35$; sharp: $B^{\rhd}=15$, $D^{\rhd}=2.56$) in (a) and
(b) for the left and the right side of the step, respectively
($C^{\lhd}=C^{\rhd}=1$ and $q^{\lhd}=q^{\rhd}=1$)\/. The
corresponding equilibrium contact angles with the substrate are indicated in the figures.
The profiles correspond to times $t=55$ (dashed lines) and 4000
(solid lines), in units of $\mu b^{\lhd}/(C^{\lhd}\sigma)$\/. From
top to bottom $a=w=15$, 9.38, and 5 for the continuous DJP and 15, 6, and
5 for the sharp DJP.  Lengths are measured in units of
$b^{\lhd}$\/.}
\label{fig3}
\end{figure} 

A typical example for the DJP at a chemical step 
is given in Fig.~\ref{fig1}\/. The DJP $\Pi(x,y)$ obtained according to Eq.~\eqref{fulldjp} is a continuous function of both $x$ and $y$.
Following Refs.~\cite{Schwartz:1998,Thiele:2006} we also consider the discontinuous
approximation
\begin{equation}
\label{pisimple}
\Pi_{dc}(x,y)=\left\{\begin{array}{rcl}
\Pi_{ch}^{\lhd}(y) & \text{for}&x<0\\
\Pi_{ch}^{\rhd}(y) & \text{for}&x>0
\end{array}\right.
\end{equation}
with $\Pi_{ch}^{\lhd/\rhd}(y)$ defined in Eq.~\eqref{dps}\/. In
the following, we denote these two forms as "continuous" and
"sharp", respectively.

The DJP induced lateral force on a two-dimensional droplet (i.e.,
the force per unit length on a liquid filament) spanning the
chemical step is given by
\begin{equation}
\label{force}
f=\int_{\Gamma}\Pi(x,y)\,n_{x}\,d{s}=f^{\lhd}+f^{\rhd},
\end{equation}
with $\Gamma$ as the liquid-vapor interface 
and $n_x = -\frac{dy}{ds}$ as the $x$-component of the outward
surface normal vector on $\Gamma$\/. Changing the integration
variable from the contour length $ds$ to $dy$ we obtain for the
forces due to the left and right hand part of the substrate,
respectively,
\begin{equation}
f^{\lhd/\rhd}=
-\int_{y_0^{\lhd}}^{y_\text{max}} \Pi_{ce}^{\lhd}(x<0)\,dy 
+\int_{y_0^{\rhd}}^{y_\text{max}} \Pi_{ce}^{\lhd}(x>0)\,dy,
\end{equation}
with the droplet apex height $y_\text{max}$\/. The first integral
is the force on the left part ($x<0$) of the droplet and the second
integral the force on the right part ($x>0$)\/. For large
droplets, i.e., for $y_\text{max}\gg y_0^{\lhd/\rhd}$, the upper
limit of integration $y_\text{max}$ can be approximated by
$\infty$\/. If, in addition,  the contact
lines are at a large distance (as compared to $y_0^{\lhd/\rhd}$) from the step, the main
contribution to $f^{\lhd/\rhd}$ stems from the part of the liquid-vapor
interface in the vicinity of the left/right contact line where
$\Pi_{ce}^{\lhd/\rhd}(x,y)\approx \Pi_{ch}^{\lhd\rhd}(y)$\/. 
Therefore for macroscopic drops one obtains 
$f^{\lhd/\rhd}=\mp\int_{y_0^{\lhd/\rhd}}^{\infty}{\Pi_{ch}^{\lhd/\rhd}(y)}
dy$ (which is independent of the droplet shape), and by using
Eq.~\eqref{cangle} one 
recovers the well known expression $f=\cos\theta_{eq}^{\rhd}-\cos\theta_{eq}^{\lhd}$ (in units of $\sigma$)
\cite{Greenspan:1978,Raphael:1988, Ondarcuhu:1991,Chaudhury:1992}\/.


The equilibrium contact angles $\theta_{eq}^{\lhd/\rhd}$, which determine the
force on macroscopic drops, are given by the integrals over
$\Pi_{ch}^{\lhd/\rhd}(y)$ from the equilibrium wetting film thickness to
\textit{infinity} (Eq.~\eqref{cangle}). The force on a nanodroplet, however, does
not depend on the values of the DJP for $y$ larger than its height.  Therefore, one can expect
that (a) the force on nanodroplets depends on the size of the droplets and (b)
that this force can be different for substrates with different DJP but with
nonetheless the same equilibrium contact angle.

In order to estimate the force on a droplet spanning a chemical
step --- an unstable configuration for which one cannot resort to
equilibrium shapes --- we assume a simple analytical expression $y(x)= y_0+a\,\left[
1-\left(x/w\right)^2\right]^{|x|^{10}+1}$ for
the droplet shape, which approximates a
parabola of basal width $2\,w$ and height $a$ which is
smoothly connected with a wetting film of thickness $y_0$\/. 
In particular for flat droplets, the main contribution to $f$ stems
from the regions in the vicinity of the contact line, i.e., for
relatively large distances from the chemical step, where
$\Pi_{ce}^{\lhd/\rhd}$ is well approximated by the DJP $\Pi^{\lhd/\rhd}_{ch}$ of a
homogeneous substrate\/. 

For a droplet of size $a=w=15$ Fig.~\ref{fig2}(a) shows $f^{\lhd/\rhd}$ as a function of
$B^{\lhd/\rhd}$ for substrates with $C^{\lhd/\rhd}=1$ and $D^{\lhd/\rhd}$ chosen
such that  $\theta^{\lhd}_{eq}=\theta^{\rhd}_{eq}=90^\circ$\/. Evidently, the droplet is
too small to justify the sharp approximation, but in both models,
$f^{\lhd}$ is monotonically decreasing. 
Although according to the above parameterization $y(x)$ the contact angles $\theta_{eq}^{\lhd/\rhd}=90^\circ$ are fixed, the force depends on the
substrate parameters such that due to $f^{\lhd}+f^{\rhd}\neq0$ a droplet on top of
a step  between two sides with equal $\theta_{eq}$ but different
chemical composition is driven to one or the other side of the
step. A macroscopic drop does not move at all in this case.
The chosen parameter values ($C^\lhd=C^\rhd$ and $q^\lhd=q^\rhd=1$)
correspond to two substrates with equal Hamaker constants.
Therefore, to leading order in the distance from the step, the long-ranged 
lateral force on a droplet sitting next to (and not on top
of) a chemical step is zero \cite{Moosavi:2008}\/. A careful
analysis of the force on a slightly displaced droplet
indicates, that the droplet stops when the trailing edge reaches
the contact line. At this point the force changes sign.

This behavior is confirmed by mesoscopic hydrodynamic calculations based on
the 2D Stokes equation for an incompressible liquid exposed to either 
the continuous or the sharp DJP (see Figs.~\ref{fig2}(c) and (d)).
This dynamics also takes into account the shape relaxation of the
droplet\/. In dimensionless form the continuity and Stokes equations read
$\bm{\nabla}\cdot\mathbf{u}=0\quad\text{and} \quad
C^{\lhd}\:\bm{\nabla}^2{\mathbf{u}}=\bm{\nabla}p$, respectively,
where $\mathbf{u}=(u_{x},u_{y})$ is the velocity vector and  $p$ 
is the pressure \cite{Moosavi:2008}\/. With the viscosity
$\mu$, we use as the velocity and the time scale $C^{\lhd}\,\sigma/\mu$ and
$\mu b^{\lhd}/(C^{\lhd}\sigma)$, respectively. Lengths and pressure
are expressed in units of $b^{\lhd}$ and $\sigma/{b^{\lhd}}$,
respectively.  We have solved theses equations numerically with a
standard biharmonic boundary integral method \cite{Kelmanson:1983}.
A no-slip boundary condition has been employed for the impermeable
liquid-solid interface and  a no-flux boundary condition was
imposed at the left and right end of the system.  At the
liquid-vapor interface the tangential stresses are assumed to be
zero and the normal stresses are balanced by surface
tension and the DJP \cite{Kelmanson:1983,Oron:1997} leading to 
\begin{equation}
{\bf n}\cdot\tau\cdot{\bf n}=-\kappa+\Pi,
\end{equation}
with the local mean curvature $\kappa$, the stress tensor $\tau$,
and the surface normal vector ${\bf n}$ pointing out of the
liquid. Here, the parameterization of the droplet
shape used in the force calculation was used as the initial configuration.

In Fig.~\ref{fig2} we show that the net force on a droplet can be 
non-zero even for equal macroscopic equilibrium contact angles on both sides. Since the contact angle as
well as the force vary monotonically with the parameters,
the net force is not necessarily directed towards
the more wettable side of the step. As an example
Figs.~\ref{fig3}(a) and (b) show various values of $B^{\lhd/\rhd}$ 
and $D^{\lhd/\rhd}$ for which the absolute value of the lateral force on
a nanodroplet (with dimensions $a=w=5$) due to left/right side of the substrate is
equal to $|f^{\lhd/\rhd}|=0.705$ or $|f^{\lhd/\rhd}|=0.765$
(continuous) and 
$|f^{\lhd/\rhd}|=0.8$ or $|f^{\lhd/\rhd}|=0.825$ (sharp)\/. As function of these products $B^{\lhd/\rhd}$ and
$D^{\lhd/\rhd}$, $\theta^{\lhd/\rhd}_{eq}$ changes accordingly. For
both values of $f^{\lhd/\rhd}$  one can find system parameters such
as the ones denoted as I and II
so that the force $f^{\lhd}_{\text{I}}$ from the less wettable side I is larger
than $f^{\rhd}_{\text{II}}$ from the more wettable side II. Consequently,
for such systems a droplet moves in the direction opposite to the
one expected on the basis of the difference of contact
angles. As in the example shown in Fig.~\ref{fig2}, the force
analysis for displaced droplets indicates that, independent of the
droplet size, the droplets will stop their motion once the
trailing contact line gets close to the chemical step. 
Both behaviors are also confirmed by the Stokes dynamics shown in
Figs.~\ref{fig3}(c) and (d): a small droplet of dimensions $a=w=5$ moves
towards the less wettable side, while a droplet of size $a=w=15$
already behaves like a macroscopic droplet in the
sense that it moves towards the more wettable side. 
For the system parameters chosen for Fig.~\ref{fig3} the critical
size, for which the direction of motion in the Stokes dynamics
changes sign, is found to be $a=w=9.5$ and
$a=w=6$ in the case of the continuous and the sharp DJP, respectively.

\begin{figure}
\center
\includegraphics[width=0.45\linewidth]{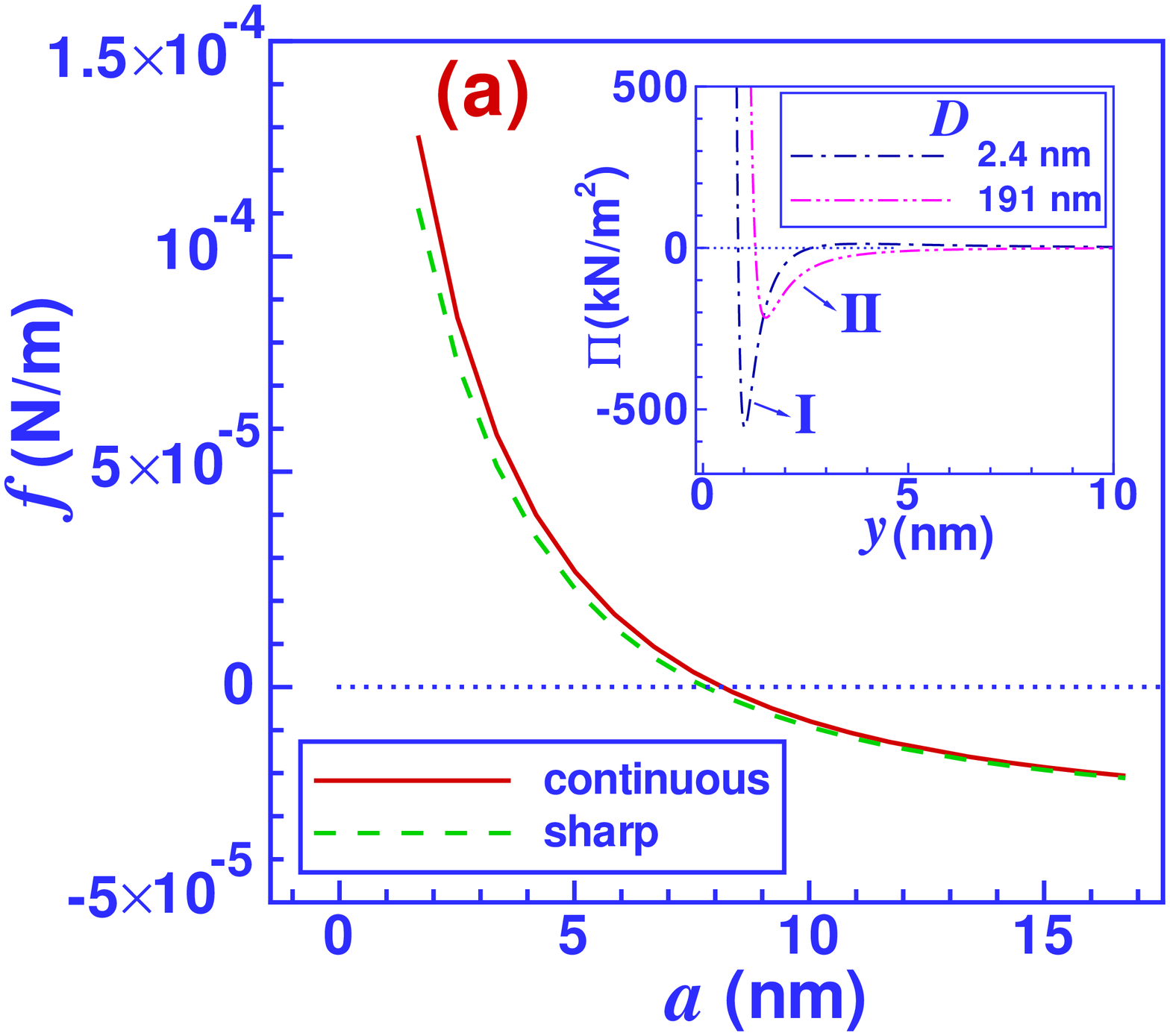}
\includegraphics[width=0.45\linewidth]{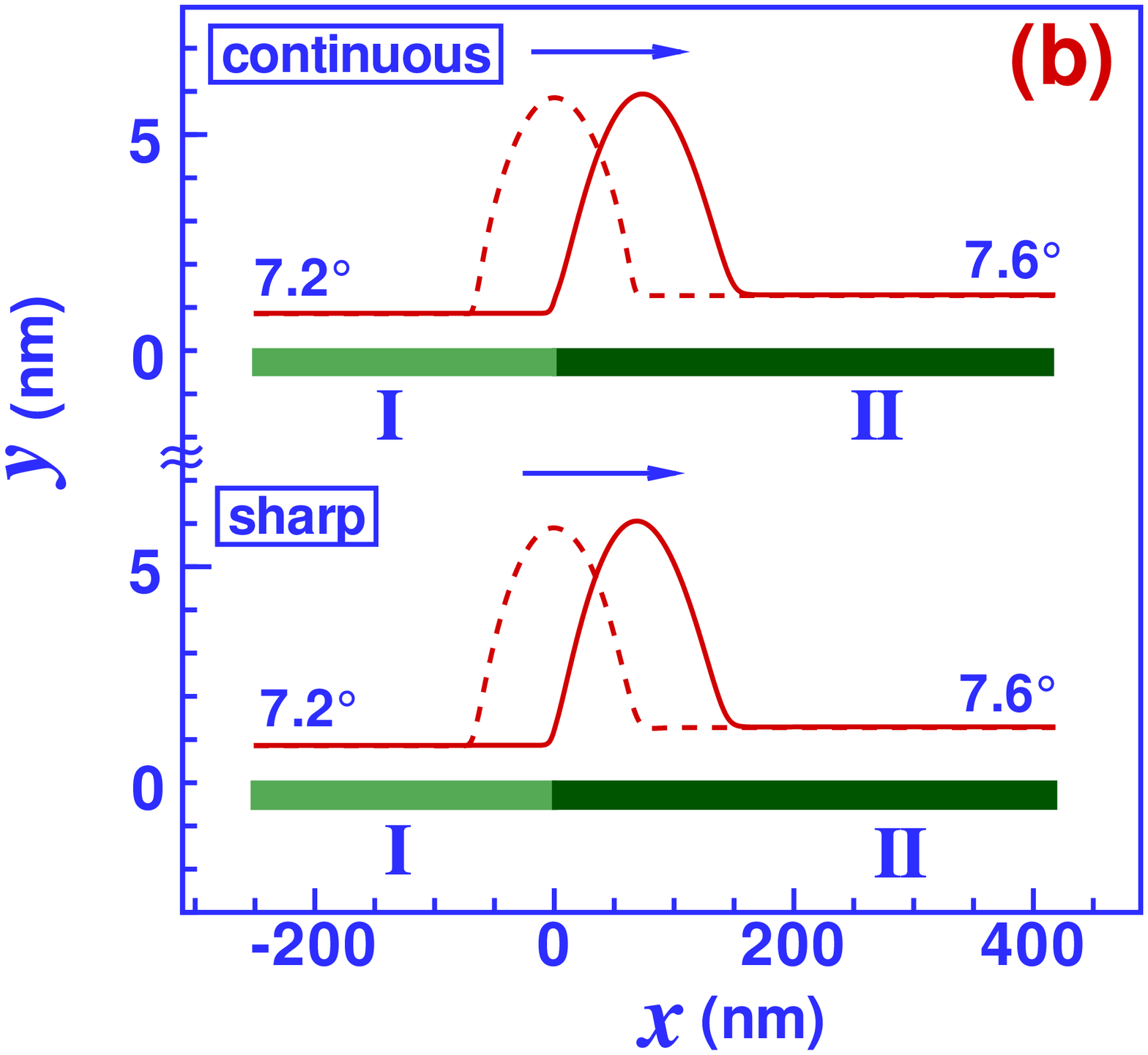}
\caption{ (a) The DJP induced force of per length on a PS droplet centered on a chemical
step composed of a Si wafer coated with oxide layers of two different
thicknesses (I: $D^{\lhd}=2.4$~nm, and II: $D^{\rhd}=191$~nm) as a function
of the droplet height $a$\/. 
The ratio between the height $a$ and the width $w$ in the ansatz for the droplet shape is chosen to be $\tan(\alpha)$ with $\alpha=7.5^\circ$, i.e., in between the equilibrium contact angle $7.2^\circ$ and $7.6^\circ$ on the left and the right hand side, respectively.
For $a\approx 7.5$ nm the force changes
sign. The inset shows the DJP on both sides. (b) Stokes dynamics of
a PS droplet with an initial height $a=$5~nm on a chemical step composed of coatings of type I
on the left side and of type II on the right side. In (b) the
profiles correspond to times $t=0.1$~s (dashed lines) and to
$t=1050$~s and 1200~s (solid lines) for the continuous (top) and
the sharp (bottom) DJP, respectively.}
\label{fig4}
\end{figure} 

Although the phenomenon described in Fig.~\ref{fig3}, i.e., the
motion towards larger contact angles, is not generic, it can be
expected to appear in the extensively studied experimental system
 polystyrene (PS: $\sigma=30.8\,\text{mN}/\text{m}$,
$\mu=1200\,\text{Pa}\,\text{s}$
\cite{Becker:2003}) on a silicon waver covered by
SiO \cite{Seemann:2001}. The experimental study reported in the Ref.~\cite{Seemann:2001}
indicates that the DJP of this system, which is shown in
the inset of Fig.~\ref{fig4}(a)
for two different thicknesses of oxide layers SiO
($2.4\,\text{nm}$ and $191\,\text{nm}$), can be represented by an expression of the
form given in Eq.~\eqref{dps}\/. 
The corresponding equilibrium contact angles are
$\theta^{\lhd}_{eq}=7.2^\circ$ ($C^{\lhd}=0.054$, $B^{\lhd}=-5.91$)
and $\theta_{eq}^{\rhd}=7.6^\circ$ ($C^{\rhd}=0.0155$,
$B^{\rhd}=-5.91$). A macroscopic droplet sitting on the chemical
step should therefore move towards the side with the thinner oxide
layer. The force calculation for droplets of varying sizes
shows, however, a change of sign at a droplet height of about $7.5\,\text{nm}$ for both the sharp and the
continous DJP. In agreement with the force
calculation, the Stokes dynamics calculations predict that a
droplet of height $a=$5~nm moves towards the less wettable substrate. In the Stokes dynamics the droplet stops once its
trailing contact line reaches the step. It covers about 80~nm in roughly 1100~s ($b^{\lhd}=0.84$~nm), which
should be experimentally observable.

In summary our study of the dynamics of droplets spanning chemical steps 
has revealed a qualitatively different behavior of
nanodroplets as compared to  macroscopic droplets
\cite{Greenspan:1978,Raphael:1988,Ondarcuhu:1991,Chaudhury:1992}\/.
While the dynamic behavior of macroscopic droplets is solely
determined by the difference of the equilibrium contact angles, the
direction of motion of nanodroplets depends on their size and
on details of the composition of the substrate. Nanodroplets 
can move towards one side of a chemical step under conditions 
for which macroscopic droplets either do not move at all or move to the 
opposite side of the step. Since most actual substrates exhibit nanoscale chemical
heterogeneities, 
our results provide a basis for a better understanding of the
dynamics of wetting phenomena, in addition to possible applications
for handling,  controlling, and guiding liquids in emerging
nanofluidic devices. 
For instance, because the size, for which the
direction of motion changes sign, also depends on the chemical composition of
the droplets,
this size selectivity has potential applications for
droplet sorting by size or by composition in nanoscale fluidic devices.

\acknowledgments{M. Rauscher acknowledges DFG support under grant number RA 1061/2-1 (SPP  1164).}


\end{document}